\documentclass[a4paper,12pt]{article}
\usepackage{amsmath,amssymb}
\usepackage{bm}
\usepackage[dvipdfmx]{color,graphicx}
\usepackage{subfigure}
\usepackage{verbatim}
\usepackage{wrapfig}
\usepackage{ascmac}
\usepackage{makeidx}
\usepackage{url}%bibTeX URL
\usepackage{fancybox}  
\usepackage{amsfonts}
\usepackage{booktabs}%tabuler
\usepackage{indentfirst}
\definecolor {gray} {gray} {0.3} %gray 1 is white
\usepackage{feynmp}
\newcommand{\Slash}[1]{{\ooalign{\hfil#1\hfil\crcr\raise.167ex\hbox{/}}}}
\newcommand{\non}{\nonumber \\ }
\newcommand{\beq}{\begin{eqnarray}}
\newcommand{\eeq}{\end{eqnarray}}
\def\thefootnote{\ifnum\c@footnote>\z@\textasteriskcentered\@arabic\c@footnote\fi}
\makeatletter
\renewcommand{\footnoterule}{%
\kern-3\p@
\hrule width 0.4\columnwidth
\kern 2.6\p@}
\def\thefootnote{\ifnum\c@footnote>\z@\@arabic\c@footnote\fi}
\makeatother
\begin{document}

\begin{center}

\hfill UT-12-30\\ 
\hfill August, 2012\\

\vspace*{0.8cm}

\Large{\bf{Vacuum Stability Constraints on  the Enhancement of the \bm{$h \rightarrow \gamma \gamma $ } Rate  in the MSSM \\ \vspace*{1.5cm}}}

\large{\textrm{\bf  Teppei Kitahara}}${}^\dag$ \footnote[0]{${}^\dag$ Electronic address: kitahara@hep-th.phys.s.u-tokyo.ac.jp } \vspace*{0.8cm}

\textit{Department of Physics, The University of Tokyo, \\  \vspace{2mm} Tokyo 113-0033, Japan} \vspace*{1.6cm}

\end{center}

%%%%%%%%%%%%%%%%%%%%%%%%%%%%%%%%%%
%%%%%%%%%%%%%%%%%%%%%%%%%%%%%%%%%%
%%%%%%%%%%%%%%%%%%%%%%%%%%%%%%%%%%
%%%%%%%%%%%%%%%%%%%%%%%%%%%%%%%%%%
%%%%%                                  Abstract                            %%%%%%
\begin{abstract}
The ATLAS and CMS collaborations discovered a new boson particle. If the new boson is the Higgs boson, the diphoton signal strength is 1.5 - 1.8 times larger than the Standard Model (SM) prediction, while the $WW$ and $ZZ$ signal strengths are in agreement with the SM one. In the Minimal Supersymmetric Standard Model (MSSM), overall consistency can be achieved by a light stau and the large left-right mixing of staus. However, a light stau and large left-right mixing of staus may suffer from vacuum instability. We first apply the vacuum meta-stability condition to the Higgs to diphoton decay rate in the MSSM. We show that the vacuum meta-stablity severely constrains  the enhancement to the Higgs to diphoton rate. For example, when the lighter stau mass is $100$ GeV, the upper bound on the enhancement to the Higgs to diphoton rate becomes  $25\%$.

\end{abstract}
{\small \hspace{21pt}  \textsc{ Keywords:} Supersymmetry, Higgs boson, Higgs to diphoton rate}
\thispagestyle{empty}
\newpage
%\begin{flushleft}
%%%%%%%%%%%%%%%%%%%%%%%%%%%%%%%%%%
%%%%%%%%%%%%%%%%%%%%%%%%%%%%%%%%%%
%%%%%%%%%%%%%%%%%%%%%%%%%%%%%%%%%%
%%%%%%%%%%%%%%%%%%%%%%%%%%%%%%%%%%
%%%%%                               Introduction                         %%%%%%
\section{Introduction}

The ATLAS  \cite{:2012gk} and CMS \cite{:2012gu} collaborations at the Large Hadron Collider (LHC) have announced the  discovery of the new boson particle around the mass region of $126$ GeV in the search for the Standard Model (SM) Higgs boson with significances of $6$ $\sigma$ (ATLAS) and $5 $ $\sigma$ (CMS), respectively. This boson has a signal strength almost consistent with the prediction of SM Higgs boson except in the diphoton channel. The signal strength $\mu (X)$ is defined by 
\begin{align}
\mu (X)  \equiv &\frac{\sigma (pp\rightarrow h) BR(h \rightarrow X)}{\sigma (pp\rightarrow h)_{SM} BR(h \rightarrow X)_{SM}} \notag \\
= & \frac{\sigma (pp\rightarrow h)}{\sigma (pp\rightarrow h)_{SM} } \times \frac{\Gamma (h \rightarrow \textrm{All})_{SM}}{\Gamma (h \rightarrow \textrm{All})}\times \frac{\Gamma (h \rightarrow X)}{\Gamma (h \rightarrow X)_{SM}} , \label{mu}
\end{align}
where  $X$ indicates a final state of the Higgs decay, for example $b \bar{b}$, $WW$ or $\gamma \gamma$.
Both the ATLAS and the CMS collaborations have reported that the observed diphoton signal strength is $1.5 - 1.8$ times larger than the SM prediction value \cite{:2012gk, :2012gu},
\beq
\mu(\gamma \gamma )_\textrm{ATLAS}& = &1.8 \pm 0.5, \non
\mu(\gamma \gamma )_\textrm{CMS} &=&1.56 \pm 0.43.\label{ob1}
\eeq
On the other hand, $\mu (ZZ)$ and $\mu (WW)$ are consistent with the SM,
\beq
\mu(ZZ^{(\ast)} \rightarrow 4 l)_\textrm{ATLAS}& = &1.4\pm 0.6, \non
\mu(ZZ^{(\ast)} \rightarrow 4 l)_\textrm{CMS} &=&0.7 { }^{+0.4}_{-0.3},\non
\mu(WW^{(\ast)} \rightarrow l \nu l \nu)_\textrm{ATLAS}& = &1.3\pm 0.5, \non
\mu(WW^{(\ast)} \rightarrow l \nu l \nu)_\textrm{CMS}& = &0.6\pm 0.4.\label{ob2}
\eeq

Although statistics of accumulated events are still low, this enhanced diphoton signal strength implies various new physics models beyond the SM \cite{Carena:2012xa, Chiang:2012qz, Cheon:2012rh, Buckley:2012em,  An:2012vp, Joglekar:2012hb,ArkaniHamed:2012kq,Haba:2012zt,Almeida:2012bq,Delgado:2012sm, Kearney:2012zi, Espinosa:2012in,Dorsner:2012pp,SchmidtHoberg:2012yy,Reece:2012gi,Davoudiasl:2012ig}. The Minimal Supersymmetric (SUSY) Standard Model (MSSM) scenarios are known to be able to enhance $\mu(\gamma \gamma )$ in the decoupling limit, where the  lighter CP-even Higgs $h$ can acquired the mass of $126$ GeV \cite{Carena:2011aa, Cao:2012fz,Carena:2012gp, Ajaib:2012eb}, and in the non-decoupling limit, where the  heavier CP-even Higgs $H$ has the mass of  $126$ GeV \cite{Christensen:2012ei, Hagiwara:2012mg, Benbrik:2012rm}. In the former scenario, a light stau and the large left-right mixing of staus can appropriately enhance $\mu(\gamma \gamma )$ \cite{Carena:2011aa,Cao:2012fz, Carena:2012gp, Ajaib:2012eb}. However, 
 it was pointed out  that the light stau and the large left-right mixing of staus  may suffer from vacuum instability \cite{CasasLleydaMunoz1996, Rattazzi:1996fb, Hisano:2010re}. For this reason, in this paper we  analyze the Higgs to diphoton rate   in a  broad parameter region in the MSSM, applying  vacuum stability conditions. We do not assume any particular high energy supersymmetry breaking structure.  In addition, we show that  vacuum stablity severely constrains  the enhancement to the Higgs to diphoton rate, and that there is an upper bound of $25 \%$ on  the enhancement to the  Higgs to diphoton rate  when the lighter stau mass is larger than $100$ GeV.

This paper is organized as follows.  In section \ref{sec2},  the enhancement to the diphoton signal strength $\mu (\gamma \gamma)$ in the MSSM will be reviewed, and the necessity for light stau and large left-right mixing of staus will be discussed. In section \ref{sec3}, we will discuss the vacuum meta-stability of staus. In section \ref{sec4}, we will analyze numerically the Higgs to diphoton rate under  the stau vacuum meta-stability condition  in a broad parameter region. 
Section \ref{sec5} is devoted to our conclusions and discussion.

%%%%%%%%%%%%%%%%%%%%%%%%%%%%%%%%%%
%%%%%%%%%%%%%%%%%%%%%%%%%%%%%%%%%%
%%%%%%%%%%%%%%%%%%%%%%%%%%%%%%%%%%
%%%%%%%%%%%%%%%%%%%%%%%%%%%%%%%%%%
%%%%%                               main plot                         %%%%%%
\section{Corrections to the signal strength from the MSSM sector}\label{sec2}

In this section, let us briefly  review the enhancement to the diphoton signal strength $\mu (\gamma \gamma)$ in the MSSM.

In general, Eq.~(\ref{mu}) implies that
there are three ways to enhance the diphoton signal strength $\mu (\gamma \gamma)$. The first way is to enhance the Higgs production cross-section $\sigma (pp\rightarrow h)$. Since the Higgs production cross-section is dominated by gluon fusion $\sigma (g g \rightarrow h)$,  this can be achieved by simply adding  new colored particles \cite{Dermisek:2007fi, Ajaib:2012eb}. The second way is to suppress the Higgs total decay width $\Gamma (h \rightarrow \textrm{All})$. Since the Higgs  mainly decays into  $b \bar{b} $, one should suppress the Higgs to $b \bar{b} $ partial width $\Gamma (h \rightarrow b \bar{b}) $. For example, in the MSSM, large  squark left-right mixing parameters, a small CP-odd Higgs mass  ($M_A$ ), and a moderate value of $\tan \beta$ ( $\langle H_d^0 \rangle =v_1= v \cos \beta,$ $\langle H_u^0 \rangle =v_2= v \sin \beta$, $\tan \beta = v_2 / v_1$ and  $v\simeq 174$ GeV) can lead to the desired suppression of the Higgs to $b \bar{b} $ partial width, so-called  ``small $\alpha_{\textrm{eff}}$ scenario'' \cite{Carena:2002qg}. Singlet multiplets extended  MSSM (e.g. NMSSM) can also lead to suppress the Higgs to $b \bar{b} $ partial width because of singlet-doublet Higgs mixing effects \cite{Ellwanger:2011aa}. The last way is to enhance the Higgs to $\gamma \gamma $ partial width $\Gamma (h \rightarrow \gamma \gamma )$ itself.
However, since these three ways are  intricately related, 
the analysis of  enhancement to the diphoton signal strength is involved.   

In the MSSM, if one takes the observations Eqs.~(\ref{ob1}) and (\ref{ob2})  into account, the  situation becomes somewhat simplified.
Eq.~(\ref{mu}) also implies that the first two ways are independent of the Higgs decay channel ($X$), and only the third pattern depends on the Higgs decay channel. 
In addition, in the MSSM, the Higgs to $WW$ and $ZZ$ partial  widths are almost equal to the SM values, since only gauge couplings contribute at leading order and SUSY particle contributions receive loop suppression. Therefore, if one employs  the first or second way to enhance the diphoton signal strength, 
additional enhancement to the $WW$ and $ZZ$ signal strengths are inevitable.
However, in fact, the observed signal strengths except for diphoton, particularly  for $WW$ and $ZZ$  are in good agreement with the SM prediction in the range of $1$ $\sigma$ \cite{:2012gk,:2012gu}.  Hence,  in the following, we assume  that  $ \sigma (pp\rightarrow h)/\sigma (pp\rightarrow h)_{SM}  \times \Gamma (h \rightarrow \textrm{All})_{SM}/\Gamma (h \rightarrow \textrm{All})$ is almost unity, and we consider only the third way. 
For that reason, we have investigated  not the diphoton signal strength $\mu (\gamma \gamma )$ but  the Higgs to diphoton partial width as compared with the SM prediction $\Gamma (h \rightarrow \gamma \gamma ) / \Gamma (h \rightarrow \gamma \gamma ) _{SM}$ in detail in this paper.

In the MSSM, the Higgs to diphoton partial width arises dominantly from the  $W$ boson loop, and sub-dominantly from the top quark loop. An analytic expression for the Higgs to diphoton partial width is given in Refs.~\cite{Shifman:1979eb, HiggsHunter} and it is  rewritten as follows,
\beq
\Gamma (h \rightarrow \gamma\gamma) = \frac{\alpha ^2 m_h^3 }{1024 \pi^3} \left | \frac{g_{hWW}} {m_W^2} A^{h}_1 (\tau _W) +\sum_{f} \frac{2 g_{hff}}{m_f} N_{c,f}Q_f^2 A^h_{\frac{1}{2} }(\tau_f) +  A^{h \gamma \gamma}_{SUSY}\right |^2,
\eeq
where $\tau_i = m_h^2 / 4 m_i^2$, $m_h$ is the lightest CP-even Higgs mass, $N_{c,i}$ is the number of colors of particle $i$, $Q_{i}$ is electric charge of particle $i$, and
\beq
A^{h \gamma \gamma}_{\text{SUSY}}=\sum_{\tilde{f}} \frac{ g_{h \tilde{f} \tilde{f}}}{ m_{\tilde{f}}^2 } N_{c, \tilde{f}} Q_{\tilde{f}}^2 A_{0}^h (\tau_{\tilde{f}}) + \sum_{i=1,2} \frac{ 2   g_{h \chi^{+}_{i} \chi^{-}_{i}}}{m_{ \chi^{\pm }_{i} }}A_{\frac{1}{2}}^h(\tau_{\chi^{\pm}_i})+\frac{  g_{h H^{+} H^{-}} }{ m_{H^{\pm}}^2} A_0^h (\tau_{H^{\pm}}),
\eeq
with loop functions $A_{i}(\tau)^h$  the Higgs coupling constants $g$   given in Appendix \ref{AppA}.

At the heavy particle loop limit $\tau_i \ll 1 $,  the loop functions $A_{i}(\tau)^h$ take the following asymptotic value,
\beq
A_1^h \rightarrow 7,\text{　　　}A_{\frac{1}{2}}^h \rightarrow -\frac{4}{3},\text{　　　}A_0^h \rightarrow - \frac{1}{3}.
\eeq 
Since the Higgs mass is $126$ GeV and the top quark mass is $173.2$ GeV \cite{Lancaster:2011wr}, the loop functions of $W$ and top quark loop are given by
\beq
A_1^h(\tau_W) = 8.36,\text{　　　}3\times \left( \frac{2}{3}\right)^2\times A_{\frac{1}{2}}^h = - 1.84.
\eeq

The Charged Higgs loop cannot sufficiently enhance the Higgs to diphoton rate since the Higgs coupling constant ($g_{h H^{+} H^{-}}$) is dominated by   gauge couplings. The Chargino loop also cannot account for sufficient enhancement  for the same reason and by $\tan \beta$ suppression \cite{Diaz:2004qt}. On the other hand, since sfermions with large left-right mixing can have a large  Higgs coupling constant ($g_{h \tilde{f} \tilde{f}}$), a light sfermion loop can sufficiently enhance the Higgs to diphoton rate. A light stop and sbottom loop, however, would not be appropriate to enhance the diphoton signal strength. These loops usually bring larger suppression of the gluon fusion rate than the enhancement of the Higgs to diphoton rate \cite{Dermisek:2007fi,Carena:2011aa}. This is the reason  that the Higgs to diphoton amplitude is dominantly constructed by a $W$ boson loop, but the $hgg$ amplitude is dominantly constructed by a top quark loop. On the other hand, a stau loop does not influence the gluon fusion since stau does not carry  color charge. Hence, in the MSSM, only a stau loop would be appropriate to enhance the diphoton signal strength.  

\begin{figure}[tbp]
\centering
\begin{eqnarray}
\nonumber
\parbox{10mm}{
\begin{fmffile}{stauloop}
\begin{fmfgraph*}(60,40)
\fmfkeep{stauloop}
\fmfleft{i1}
\fmflabel{\Large{$h$}}{i1}
\fmfright{o1,o2}
\fmflabel{\Large{$\gamma$}}{o1}
\fmflabel{\Large{$\gamma$}}{o2}
\fmf{dashes}{i1,v1}
\fmf{dashes,left,tension=0.3,tag=1,label=\Large{$\tilde{\tau}_L$},l.side=left}{v1,v}
\fmf{dashes,left,tension=0.3,tag=2,label=\Large{$\tilde{\tau}_R$},l.side=left}{v,v1}
\fmfv{decor.shape=cross,decor.size=10}{v}
\fmf{phantom}{v,o11}
\fmf{phantom}{v,o21}
\fmf{wiggly}{o11,o1}
\fmf{wiggly}{o21,o2}
\end{fmfgraph*}
\end{fmffile} }
\textrm{ \hspace{110pt} }
\end{eqnarray}
\caption{Feynman diagram which gives rise to an enhancement to the Higgs to diphoton rate in the mass insertion method. The cross represents the left-right mixing of staus.}
\label{stauloop}
\end{figure}
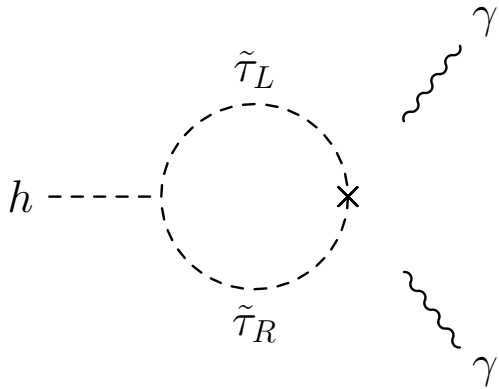

The stau loop correction to the Higgs to diphoton amplitude  is roughly proportional to $| m_{\tau } \mu \tan  \beta| / m_{\tilde{\tau}_1}^2$ in the large $\mu \tan \beta $ region, where $m_{\tilde{\tau}_1}$ is the lighter stau mass.
The corresponding Feynman diagram is shown in Figure \ref{stauloop}. Light stau and large left-right mixing, i.e. large $\mu \tan \beta$, can sufficiently enhance the Higgs to diphoton rate, so-called ``light stau scenario" \cite{Carena:2011aa, Carena:2012gp, Hagiwara:2012mg, Ajaib:2012eb}. We obtain a simple formula for the leading stau corrections as follows,
\beq
\frac{\Gamma (h \rightarrow \gamma \gamma )}{\Gamma (h \rightarrow \gamma \gamma )_{SM}} \simeq \left(1+ \sum_{i=1,2}0.05 \frac{m_\tau \mu \tan \beta }{ m_{\tilde{\tau}_i}^2}  x^{\tau _i}_L  x^{\tau _i}_R  \right)^2 ,
\eeq
where stau mass eigenstates are given by $\tilde{\tau}_i = x^{\tau_i}_L  \tilde{\tau}_L + x^{\tau_i}_R  \tilde{\tau}_R$, $(x^{\tau_i}_L)^2 + (x^{\tau_i}_R)^2=1$. 

Furthermore, the light stau scenario is also motivated by the anomalous magnetic moment of the muon \cite{Carena:2012gp, Giudice:2012pf}, where a  $3.2$ $\sigma$ discrepancy between the experimentally measured value $a_{\mu}^{\textrm{exp}} $ and the theoretical prediction value in the SM $a_{\mu}^{\textrm{SM}}$, $\Delta a_{\mu} = a_{\mu}^{\textrm{exp}} -a_{\mu}^{\textrm{SM}} = ( 26.1 \pm 8.0) \times 10^{-10}  $ \cite{Hagiwara:2011af} is observed. The reason for this is that light stau is compatible light $\mu$-slepton when one considers high energy physics, and light $\mu$-slepton and large $\tan \beta$ are favored to explain the discrepancy of the anomalous magnetic moment of the muon \cite{Moroi:1995yh}.

%\newpage
%%%%%%%%%%%%%%%%%%%%%%%
%%%%%%%%%%%%%%%%%%%%%%%

\section{The Vacuum meta-stability constraints}\label{sec3}
In the MSSM, the large left-right mixing of staus  can enhance the Higgs to diphoton    rate as compared to the SM prediction. However, it is known that large left-right mixing and thus  large $\mu \tan \beta$ may suffer from vacuum instability \cite{CasasLleydaMunoz1996, Rattazzi:1996fb}. The scalar potential develops the new charge-breaking minimum which leads to the $\tilde{L} \neq 0 $ or $\tilde{\tau}_R \neq 0 $ vacuum for sufficiently large $\mu \tan \beta$. And then, the minimum becomes lower than the ordinary electroweak-breaking minimum which leads to the $\tilde{L} = 0 $ and $\tilde{\tau}_R = 0 $ and $v \neq 0$ vacuum. In order to prohibit vacuum decay to charged-breaking vacuum, the lifetime of the electroweak-breaking vacuum is required to be longer than the age of the universe.

The vacuum transition rate from the false vacuum to the true vacuum can be evaluated by semiclassical technique \cite{Coleman:1977py}. Then,
the imaginary part of the energy of the false vacuum   determines the vacuum transition rate to the true vacuum.
 In the semiclassical technique, one evaluates  the energy of the false vacuum state using the path integral method in Euclidean space-time.
The vacuum transition rate per unit volume is evaluated as follows,
\beq
\frac{\Gamma}{V} = A e^{-B},
\eeq
where a precise value of the coefficient $A$ is difficult to evaluate. However, it does not depend dramatically  on the parameters of the theory, and one can roughly estimate it at the fourth power of the typical scale in the potential,
\beq
A\simeq (100 \textrm{ GeV})^4.
\eeq 
In contrast, the power index $B$ is a sensitive parameter of the vacuum transition rate per unit volume. It can be evaluated by an $O(4)$ symmetric solution as follows,
\beq
B = S_{E} [\bar{\phi}(\rho )]- S_{E}[\phi ^f] ,
\eeq
where  $\rho$ is a radial coordinate in four-dimensional spacetime, $S_{E}[\phi]$ is the Euclidean action as follows,
\beq
 S_{E} [\phi(\rho )]=\int_{0}^{\infty} 2 \pi^2 \rho^3 d \rho \left[ \frac{1}{2} \left( \frac{d \phi}{d \rho} \right)^2 + V (\phi ) \right],
\eeq
 $\phi ^f$ is the value of the fields at false vacuum and $\bar{\phi}$ is  the bounce configuration. The bounce configuration is a stationary point of the action and also satisfies the following boundary condition,
\beq
\lim_{\rho \to  \infty} \bar{\phi}(\rho) = \phi ^f, \textrm{       \ \ \ \  } 
\left. \frac{d \bar{\phi}(\rho)}{d \rho} \right|_{\rho =0} =0.
\eeq

On the other hand, the current value of the Hubble parameter given by $H_0 \simeq 1.5 \times 10^{-42}$ GeV. When the vacuum transition rate per unit volume $\Gamma / V$ is smaller than the fourth power of $H_0$, i.e. the lifetime of false vacuum is longer than the age of the universe, the power index $B$ is larger than $403.6$. Therefore, the vacuum meta-stability condition is approximately given as follows,
\beq
B \gtrsim  400.
\eeq

A first study of the meta-stability condition of the stau sector has been conducted in Ref.~\cite{Hisano:2010re}, where the bounce configuration in three-field space (up-type neutral Higgs field and left- and right-handed stau field) was   evaluated numerically. This study was done at tree level which only includes the dominant top/stop loop correction. The scalar potential, expanded around the electroweak-breaking vacuum in three-field space is given as  follows,
\beq
V&=& \frac{1}{2} m_Z^2 \sin ^2\beta ( 1 + \Delta_t )  \phi ^2 +(m_{\tilde{L}}^2 + \frac{g^2 -g^{\prime 2}}{4} v_2^2 ) \tilde{L}^2 + ( m_{\tilde{\tau}_R}^2 + \frac{g^{\prime 2}}{2} v_2^2) \tilde{\tau}_R^2 \non
& &-2 y_{\tau} \mu \tilde{L} \tilde{\tau}_R  (v_2 + \frac{\phi}{\sqrt{2}})  +\frac{g^2 - g^{\prime 2}}{2\sqrt{2}} v_2 \phi \tilde{L}^2 + \frac{g^{\prime 2}}{\sqrt{2}} v_2 \phi \tilde{\tau}_R^2  +  \frac{ m_Z^2 \sin ^2\beta ( 1 + \Delta_t ) }{2 \sqrt{2} v_2} \phi ^3 \non 
& &+\frac{m_Z^2  ( 1 + \Delta_t )}{16 v^2} \phi^4 +\frac{g^2 + g^{\prime 2}}{8} \tilde{L}^4 + \frac{g^{\prime 2}}{2} \tilde{\tau}_R^2 +(y_{\tau}^2 - \frac{1}{2} g^{\prime 2})\tilde{L}^2 \tilde{\tau}_R^2 +\frac{g^2 - g^{\prime 2}}{8} \phi^2 \tilde{L}^2 \non 
& & + \frac{g^{\prime 2}}{4} \phi^2 \tilde{\tau}_R^2, \label{potential}
\eeq
where $H_u^0 = v_2 + \phi /\sqrt{2}$ and $\Delta_t$ is  the leading log term of the one loop corrections for top/stop loops,
\beq
\Delta_t = \frac{3}{2 \pi^2}\frac{y_t^4}{g^2 + g^{\prime 2}}\log{\frac{\sqrt{m_{\tilde{t}_1}^2 m_{\tilde{t}_2}^2}}{m_t^2}}.\label{kore}
\eeq
When the Higgs boson mass accomplishes $126$ GeV, $\Delta_t$ has to be about $1$. 
 The scalar potential Eq.~(\ref{potential})  includes only real parts of scalar bosons. 
Note that because  the tau Yukawa coupling is $y_{\tau} = m_{\tau} / v_1$, the first term in the second line of Eq.~(\ref{potential}) is proportional to $\mu \tan \beta$ for large $\tan\beta$. Therefore, this term can make a new minimum point of the scalar potential and thus has  negative influence on vacuum stability.

The obtained approximate  meta-stability conditional function \cite{Hisano:2010re} is as follows,
\beq
|\mu \tan \beta | < 76.9 \sqrt{m_{\tilde{L}} m_{\tilde{\tau }_R} } + 38.7 (m_{\tilde{L}}+ m_{\tilde{\tau }_R}) -1.04 \times 10^4 \textrm{ GeV}. \label{hisano}
\eeq
Note that the vacuum meta-stability condition is sensitive  only to $m_{\tilde{L}}^2$, $m_{\tilde{\tau}_R}^2$ and $\mu \tan\beta$, but not $\tan \beta$ or $\mu$ itself, and neither $\Delta_t$ nor $A_{\tau}$. 
In Ref.~\cite{Hisano:2010re}, it is assumed that the $H_u^0$ mass term, $\mu^2 + m_{H_u}^2$, is negative in order to get a vacuum expectation value $v_2$.
However, it is known that even when  $\mu^2 + m_{H_u}^2$ is positive, correct electroweak symmetry breaking can be accomplished  because when the $H_d^0$ component is integrated out, the $H_u^0$ mass term can become effectively negative \cite{Csaki:2008sr}.
Therefore, whether   $\mu^2 + m_{H_u}^2$ is negative or not, the vacuum condition does not change.
In order to check this,
we analyzed the shape and the global minimum of  the scalar potential and probed the upper bound of $\mu$ requiring that the electroweak-breaking minimum is the global minimum (see Appendix \ref{AppB}).
As a result, we found that regardless of the  sign of $m_{H_u}^2$, the parameter $m_{H_u}^2$ does not affect the upper bound of $\mu$.

We applied this vacuum meta-stability condition to  the Higgs to diphoton rate.  This condition means that large $\mu \tan \beta$ and small stau mass are severely restricted. As a result, the enhancement to the Higgs to diphoton rate by light stau loop is also severely restricted.

%%%%%%%%%%%%%%%%%%%%%%%
%%%%%%%%%%%%%%%%%%%%%%%

\section{Numerical analysis}\label{sec4}
In this section, we analyze numerically the Higgs to diphoton partial width as compared with the SM prediction $\Gamma (h \rightarrow \gamma \gamma ) / \Gamma (h \rightarrow \gamma \gamma ) _{SM}$ and apply the stau vacuum meta-stability condition to the  $\Gamma (h \rightarrow \gamma \gamma ) / \Gamma (h \rightarrow \gamma \gamma ) _{SM}$ in a broad parameter region. 

The lower bound of the stau mass is obtained by various collider experiments \cite{Beringer:1900zz},
\beq
m_{\tilde{\tau}1} > 81.9 \textrm{ GeV},
\eeq
and we naively adopt the lower mass bound $m_{\tilde{\tau}1} \geq 100$ GeV in the following calculation.

\begin{figure}[tbp]
 \begin{center}
  \includegraphics[width=77mm]{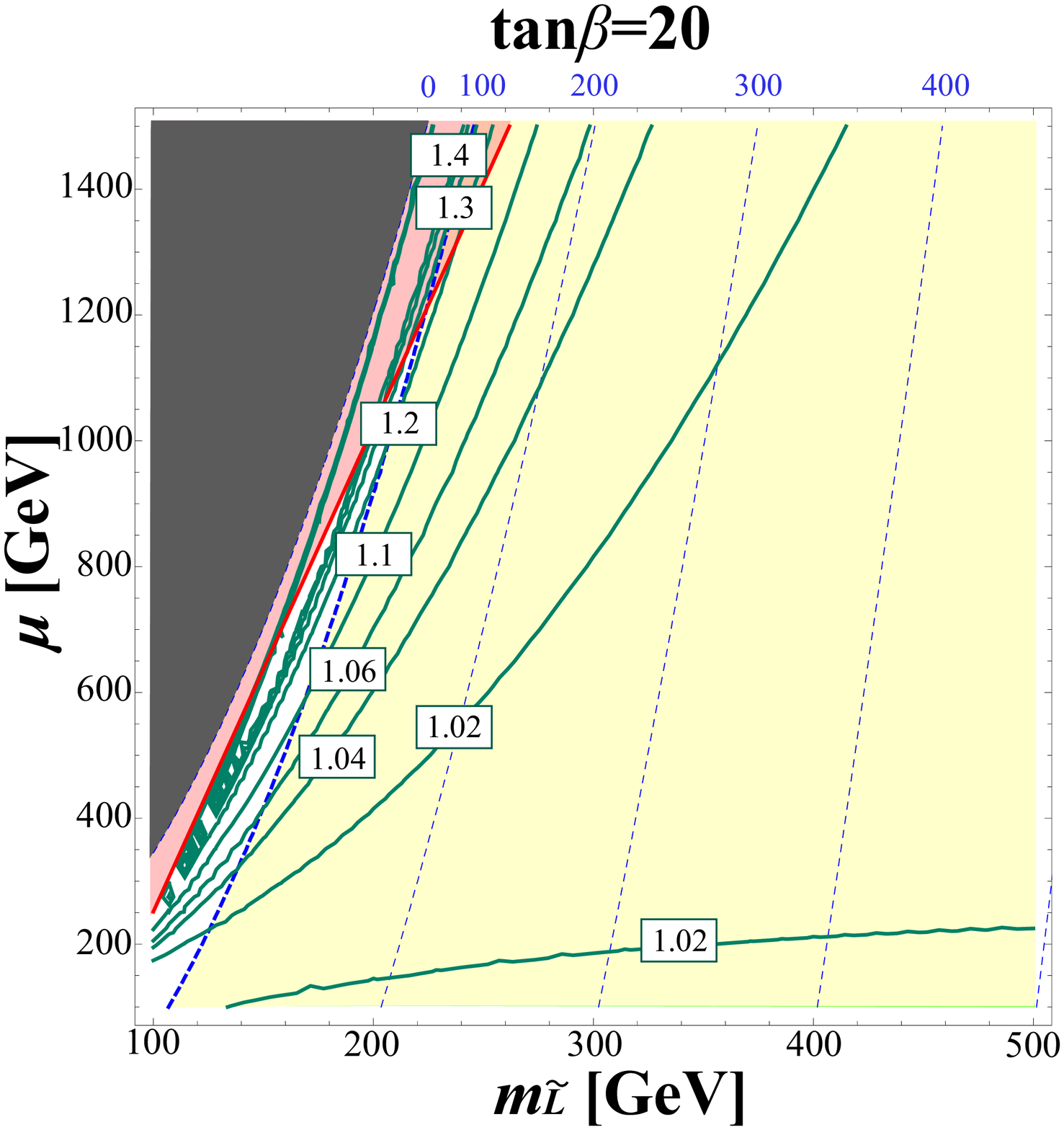}  \textrm{}  \includegraphics[width=74mm]{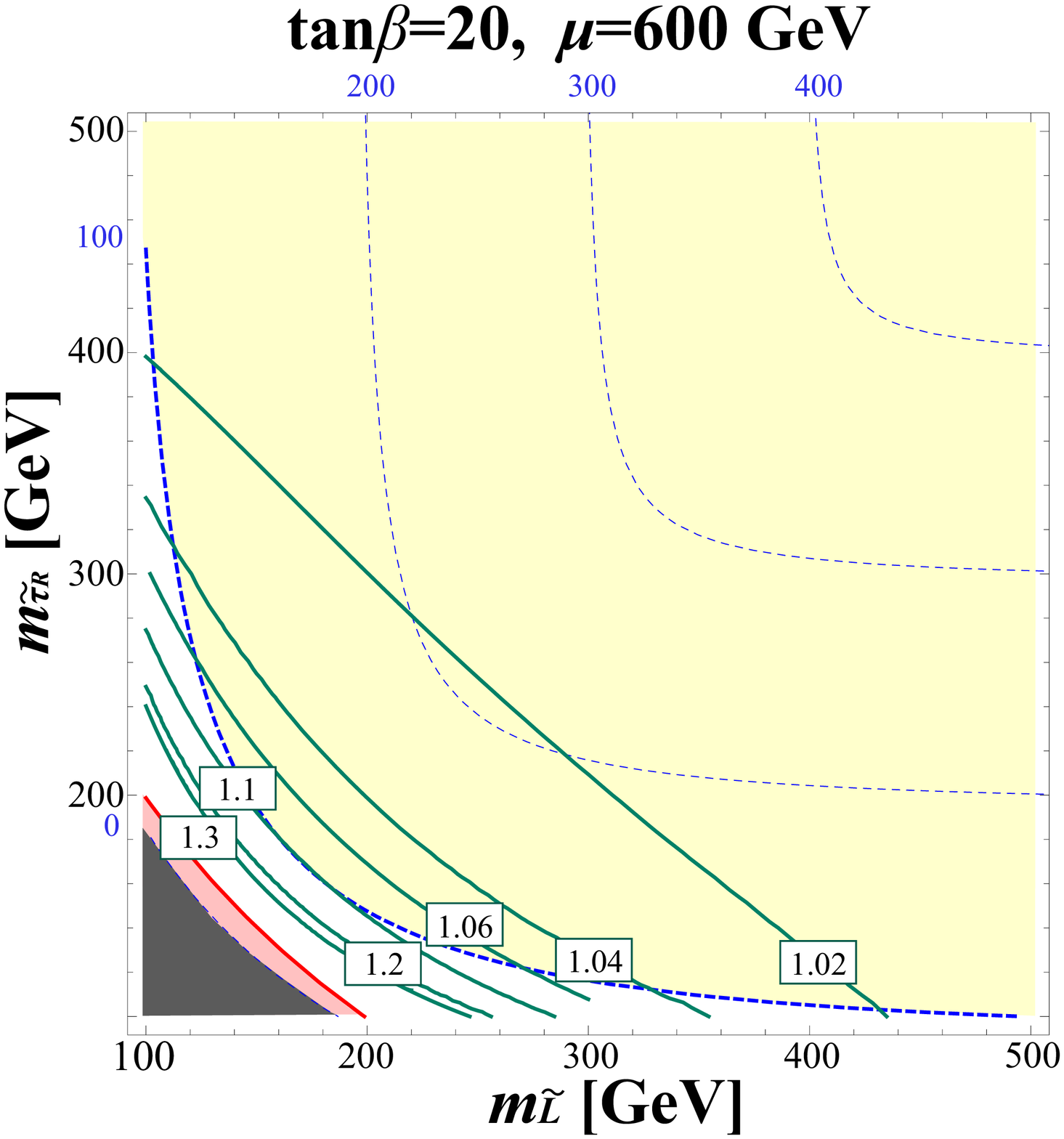} \\ \textrm{　} \\
    \includegraphics[width=77mm]{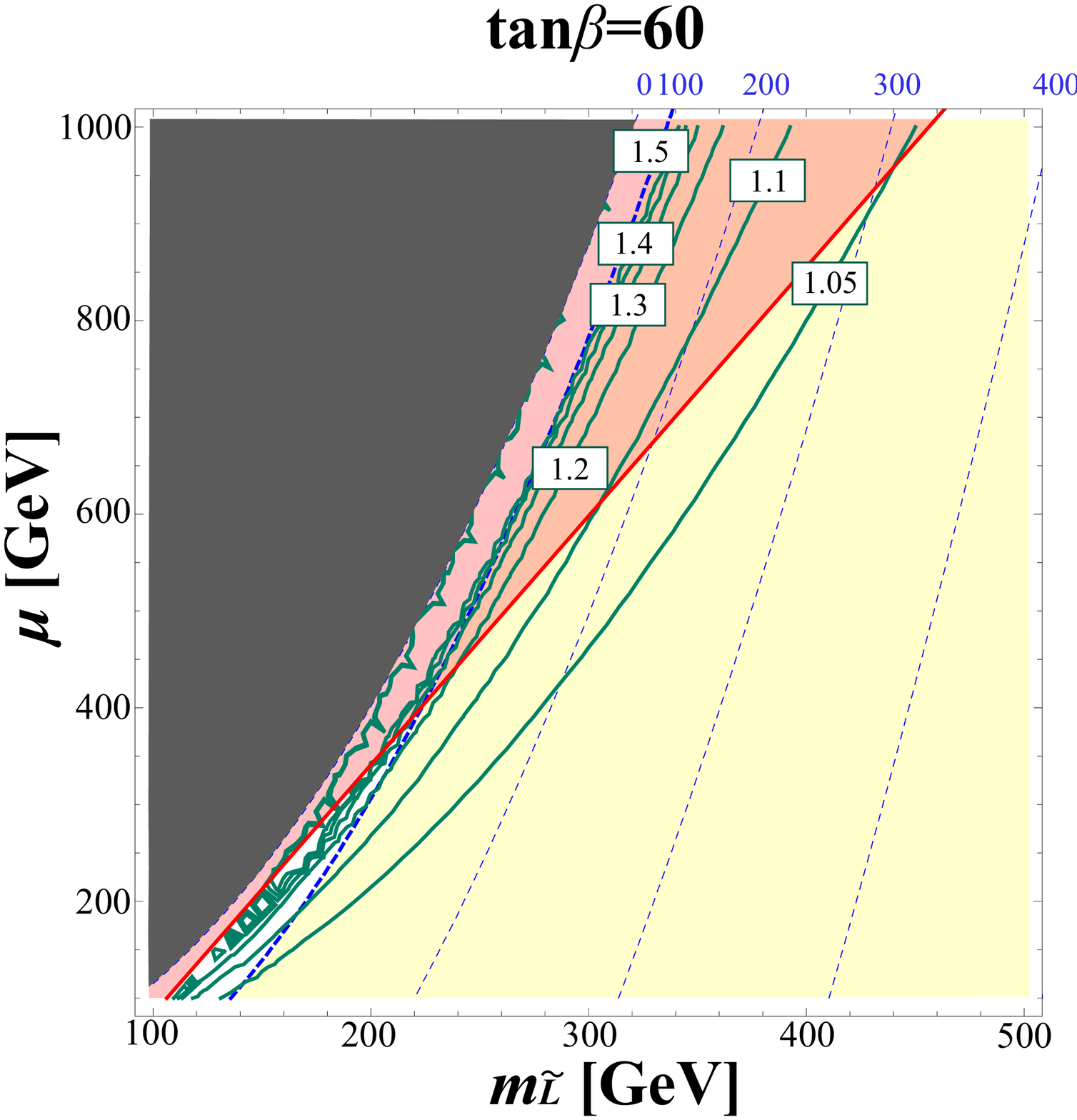}   \textrm{}  \includegraphics[width=74mm]{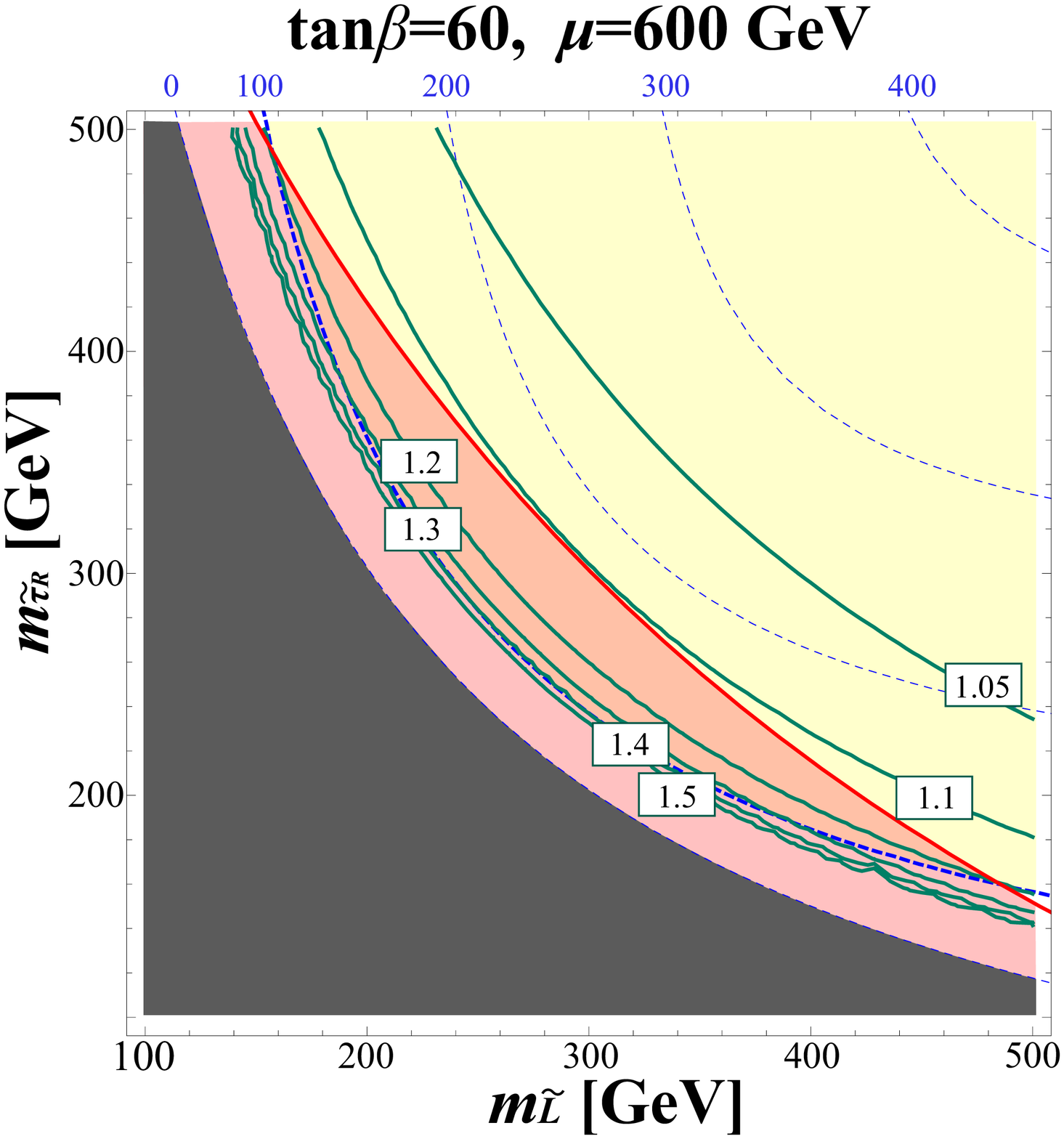}
 \end{center}
 \caption{Solid lines are contour plots of  $\Gamma (h \rightarrow \gamma \gamma ) / \Gamma (h \rightarrow \gamma \gamma )_{\text{SM}}$, in the $\mu -m_{\tilde{L}}$ plane for $m_{\tilde{\tau}R} = m_{\tilde{L}} $ (\textbf{left side}), as well as in the $m_{\tilde{L}} - m_{\tilde{\tau}R}$ plane for $\mu = 600$ GeV (\textbf{right side}). $\tan \beta = 20$ (\textbf{top}) and  $\tan \beta = 60$ (\textbf{bottom}). In all panels, $M_A = 1$ TeV, $A_{\tau} = 0$ GeV, $m_{\tilde{Q}3} = m_{\tilde{t}R} = 850$ GeV and $A_{t} = 1.7$ TeV giving $m_h \sim 126$ GeV at the two loop level. Dashed lines are contour plot of the lighter stau mass $m_{\tilde{\tau}1}$. The red  (gray)  areas are breaking the vacuum meta-stability condition. The gray (dark gray)  areas are the regions where the lighter stau is tachyonic.}
 \label{fig:1}
\end{figure}

In Figure \ref{fig:1},   the dependence of  $\Gamma (h \rightarrow \gamma \gamma ) / \Gamma (h \rightarrow \gamma \gamma )_{\text{SM}}$ in the $\mu -m_{\tilde{L}}$ plane for $m_{\tilde{\tau}R} = m_{\tilde{L}} $ (left-hand side), as well as in the $m_{\tilde{L}} - m_{\tilde{\tau}R}$ plane for $\mu = 600$ GeV (right-hand side) are shown. Solid lines represent contours of  $\Gamma (h \rightarrow \gamma \gamma ) / \Gamma (h \rightarrow \gamma \gamma )_{\text{SM}}$. Dashed lines represent contours of the lighter stau mass $m_{\tilde{\tau}1}$. The gray  (dark gray)  areas are the regions where the lighter stau is tachyonic.  The yellow (mostly white) areas are $m_{\tilde{\tau}1} \geq 100$~GeV. In addition, the red (gray) areas are the regions where the vacuum meta-stability condition  (\ref{hisano}) is broken. We considered fixed values of  $\tan \beta = 20$ (top panel) and  $\tan \beta = 60$ (bottom panel).  In all panels, $M_A = 1$~TeV, $M_3 =1$~TeV, $M_2 = 300$~GeV, $A_{\tau} = 0$~GeV, $m_{\tilde{Q}3} = m_{\tilde{t}R} =m_{\tilde{b}R}= 850$~GeV, other squark masses are $1 $~TeV and $A_{t} = 1.7$~TeV, which gives $m_h \sim 126$ GeV at the two loop level. In our  calculation, the value of $\Gamma (h \rightarrow \gamma \gamma )$ contains not only the stau loop but the full one loop order. We have checked that the results are only sensitive to  $m_{\tilde{L}}$, $m_{\tilde{\tau}R}$, $\mu$ and $\tan\beta$.
On the other hand, however, we have checked that they are insensitive to the Higgs mass. 

For $\tan\beta = 20$,  the enhancement to  $\Gamma (h \rightarrow \gamma \gamma ) / \Gamma (h \rightarrow \gamma \gamma )_{\text{SM}}$ is small in the low $\mu $ region. In the  high $\mu$ region, however, the enhancement becomes gradually  large along  light stau region. 
 The vacuum stability condition (\ref{hisano}) has given weak constraint to $\Gamma (h \rightarrow \gamma \gamma ) / \Gamma (h \rightarrow \gamma \gamma )_{\text{SM}}$. In fact, when $\mu = 600$ GeV, there are no constraints from the vacuum stability in the  $m_{\tilde{\tau}1} \geq 100$ GeV region. Here, the case of $m_{\tilde{L}} = m_{\tilde{\tau}R} \sim 170$~GeV gives a maximum enhancement of the diphoton rate around 10~$\%$.
On the other hand, for $\tan\beta=60$, the enhancement of  $\Gamma (h \rightarrow \gamma \gamma ) / \Gamma (h \rightarrow \gamma \gamma )_{\text{SM}}$ is larger than for $\tan\beta=20$. However, the enhancement receives more severe constraint from vacuum stability. As can be seen from the lower left and right panels of Figure \ref{fig:1}, large enhancement arias are violating the vacuum stability. After all,  for $\tan\beta=60$, the enhancement of the  diphoton rate only reaches around 20~$\%$.

\begin{figure}[tbp]
 \begin{center}
 \includegraphics[width=110mm]{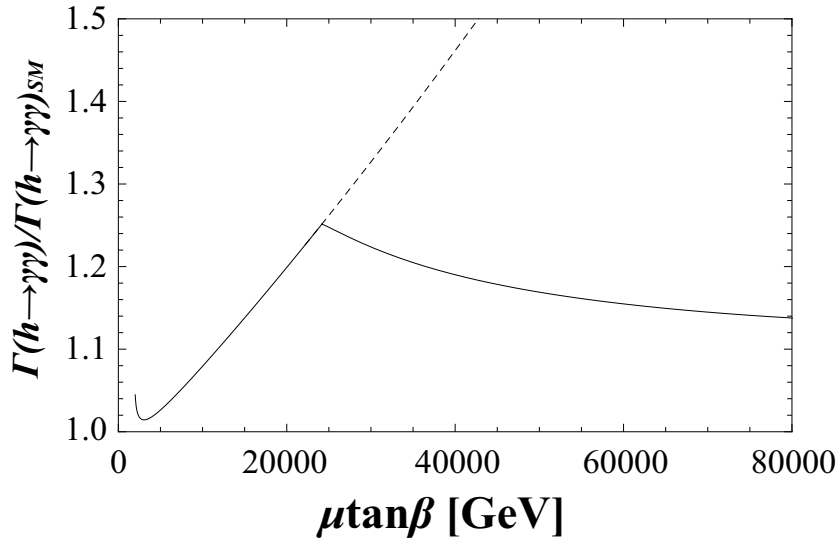}
 \end{center}
 \caption{The upper bound line of  $\Gamma (h \rightarrow \gamma \gamma ) / \Gamma (h \rightarrow \gamma \gamma )_{\text{SM}}$ as a function of $\mu \tan\beta$, for $\tan \beta = 50$, varying $m_{\tilde{L}}$ and $m_{\tilde{\tau}R} $ with remaining the lighter stau mass $m_{\tilde{\tau}1} = 100$~GeV. Other parameters  take the same value as  Figure \ref{fig:1}.  If we would not consider the vacuum stability, the upper bound of  $\Gamma (h \rightarrow \gamma \gamma ) / \Gamma (h \rightarrow \gamma \gamma )_{\text{SM}}$ is given by the dashed line.}
 \label{fig:2}
\end{figure}

Next, we show the upper bound line of  $\Gamma (h \rightarrow \gamma \gamma ) / \Gamma (h \rightarrow \gamma \gamma )_{\text{SM}}$ as a function of $\mu \tan\beta$, varying $m_{\tilde{L}}$ and $m_{\tilde{\tau}R} $ while remaining the lighter stau mass $m_{\tilde{\tau}1} = 100$ GeV, see Figure \ref{fig:2}. We fixed $\tan\beta = 50$, and all other  parameters  take  the same values as Figure~\ref{fig:1}. In the low $\mu \tan\beta $ region, $\mu \tan\beta \lesssim 24$ TeV, there are no constraints from  vacuum stability for $m_{\tilde{\tau}1} \geq 100$~GeV,  like these encountered in the upper right
 panels of Figure \ref{fig:1}. Note that the peak at very low $\mu \tan \beta $  is caused by the light chargino loop. On the other hand, in the high $\mu \tan \beta$ region,  $ 24$ TeV $ \lesssim \mu \tan\beta$, the vacuum stability condition line and the $m_{\tilde{\tau}1} =100$ GeV line  begin to cross like in the lower right
 panels of Figure \ref{fig:1}. Hence,    vacuum stability severely constrains   $\Gamma (h \rightarrow \gamma \gamma ) / \Gamma (h \rightarrow \gamma \gamma )_{\text{SM}}$. If we would not consider the vacuum stability, the upper bound line of  $\Gamma (h \rightarrow \gamma \gamma ) / \Gamma (h \rightarrow \gamma \gamma )_{\text{SM}}$ would be given by  the dashed line. 
Note that although  in Figure \ref{fig:2} we fixed $\tan \beta = 50$, we have checked that  these  results  are in fact insensitive to $\tan \beta $ ,  but  sensitive to $\mu \tan \beta$. The  reason is that the stau mass matrix,  the vacuum stability condition (\ref{hisano}) and the ratio  $\Gamma (h \rightarrow \gamma \gamma ) / \Gamma (h \rightarrow \gamma \gamma )_{\text{SM}}$ are dependent on the form of $\mu \tan \beta$  in the large $\mu \tan\beta$ region. Thus we found that  the Higgs to diphoton decay 
rate can increase only by  $25 \%$ compared to SM value, at $\mu \tan \beta \sim 24$ TeV in the MSSM.

\begin{figure}[htbp]
 \begin{center}
 \includegraphics[width=74mm]{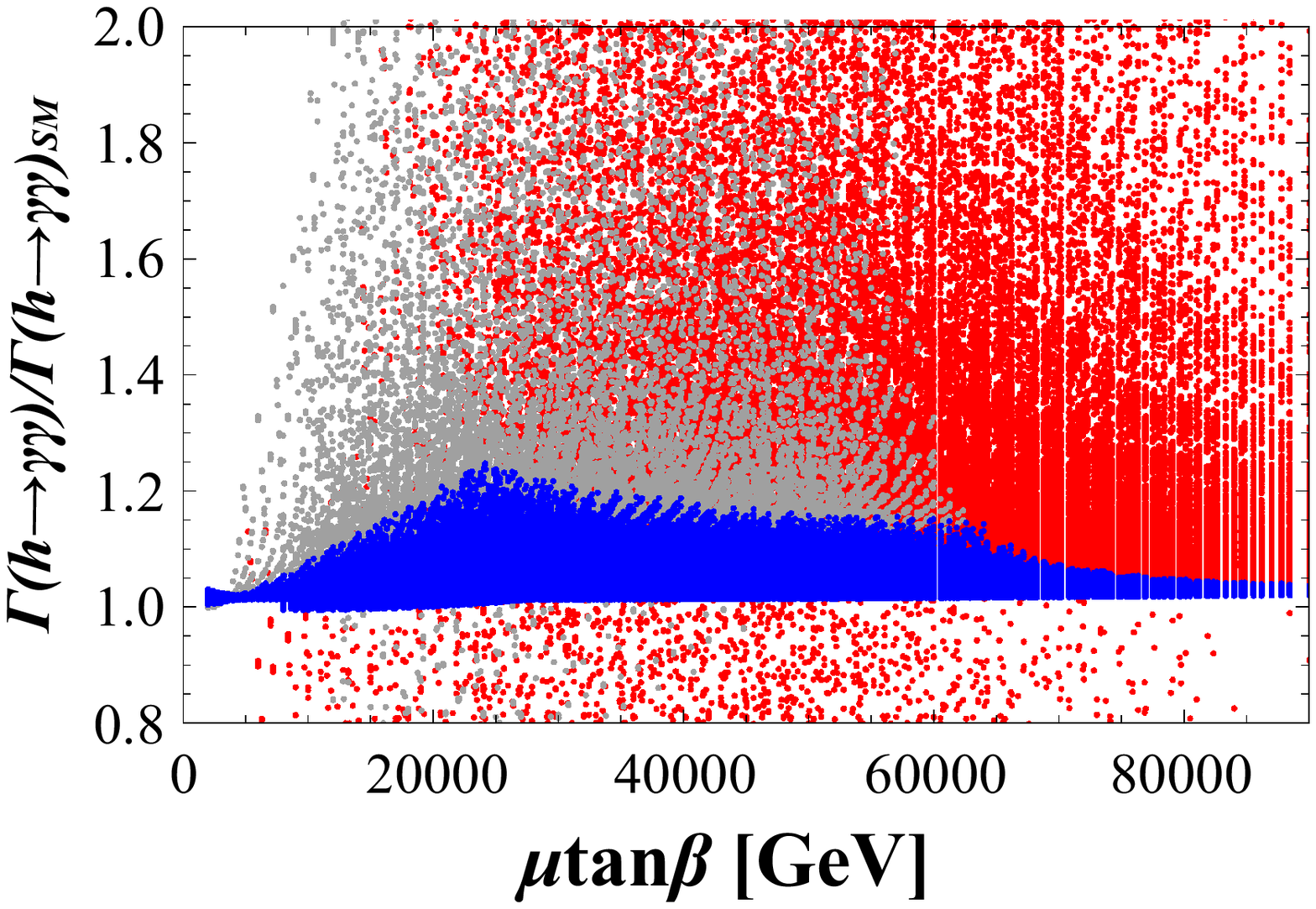} \textrm{} \includegraphics[width=77mm]{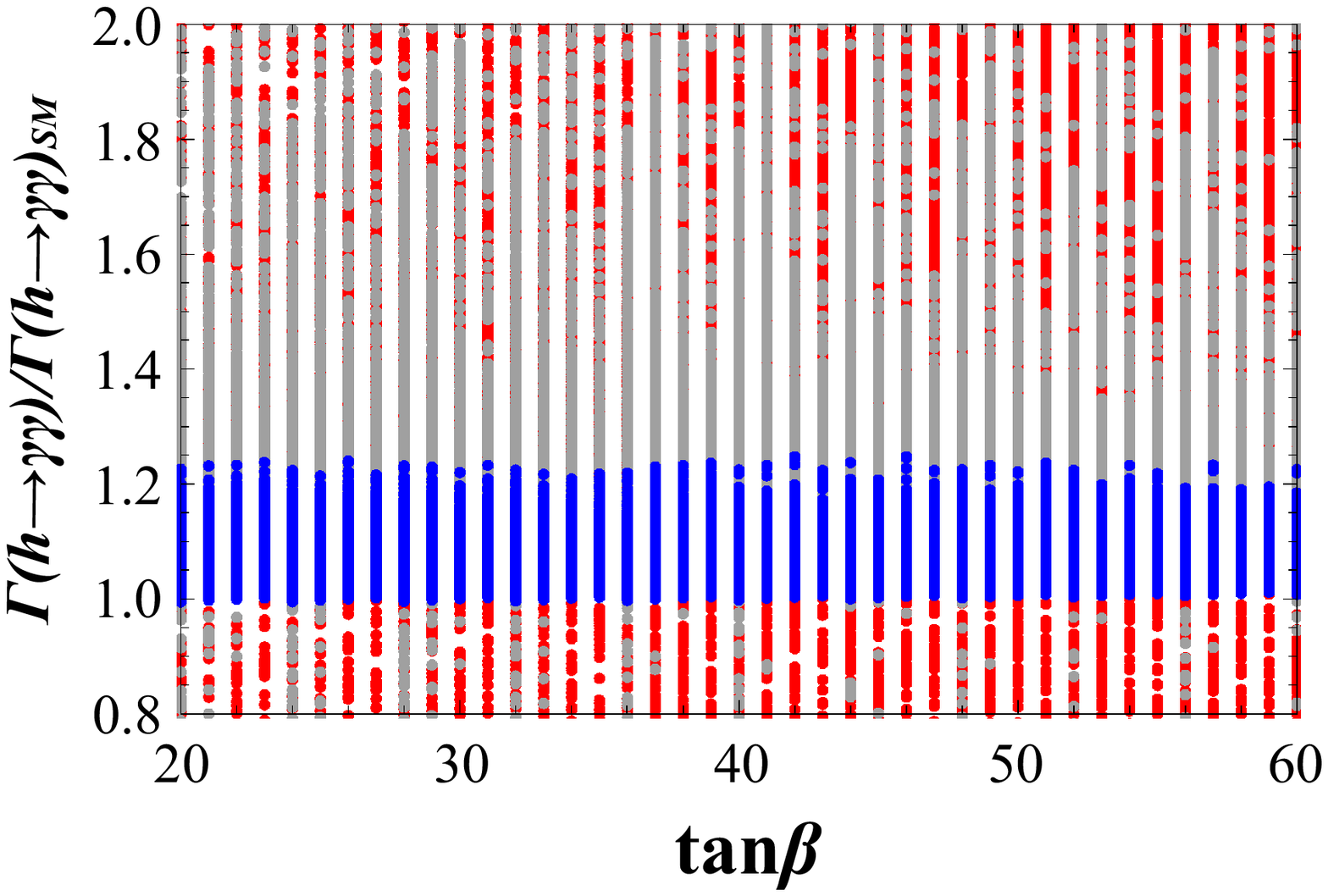} \\
  \textrm{} \\
 \includegraphics[width=100mm]{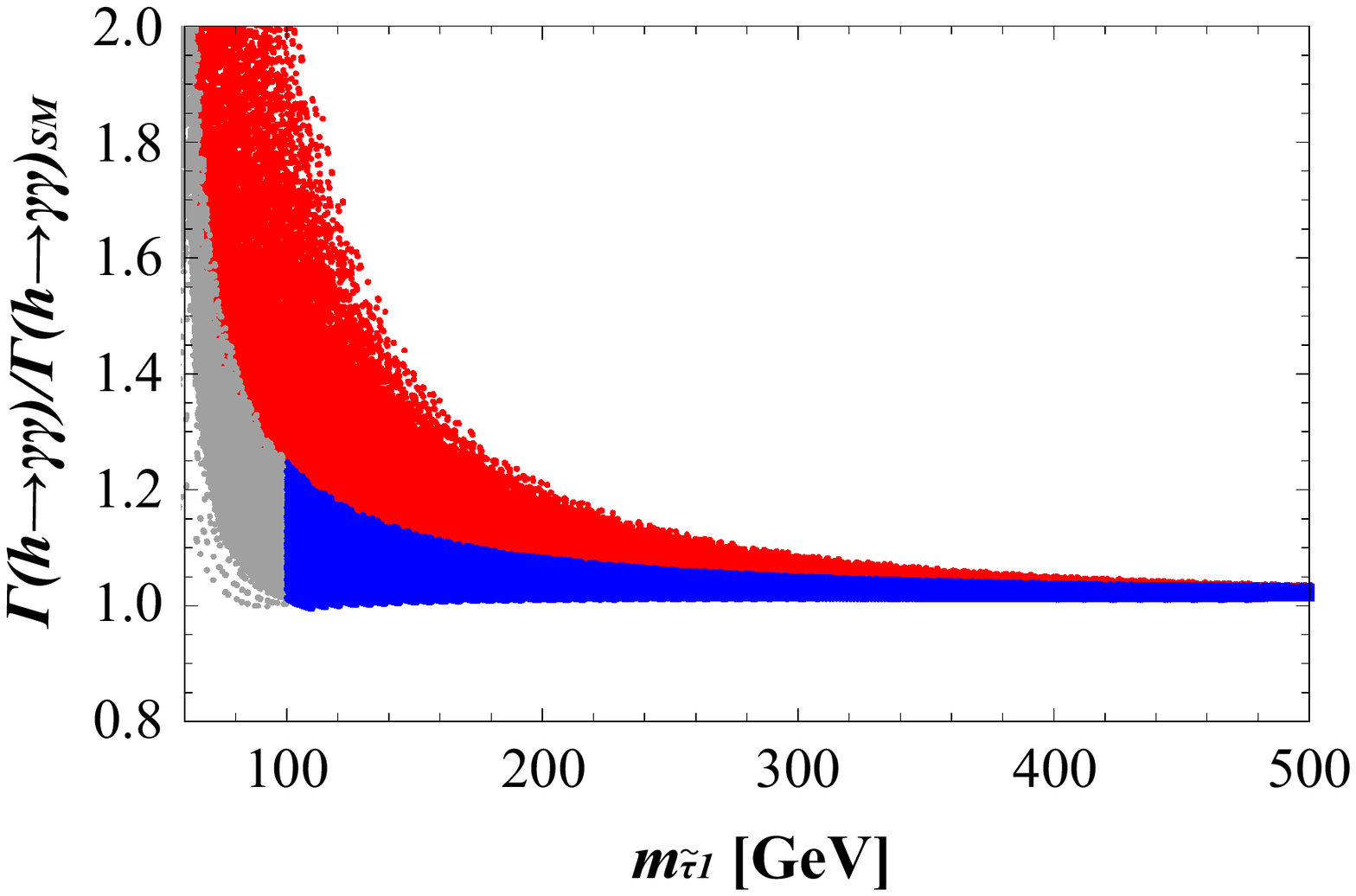}
 \end{center}
 \caption{The scatter plots of  $\Gamma (h \rightarrow \gamma \gamma ) / \Gamma (h \rightarrow \gamma \gamma )_{\text{SM}}$ as a function of $\mu \tan\beta$ (\textbf{left}), $\tan \beta$ (\textbf{right}) and the lighter stau mass $m_{\tilde{\tau}1}$ (\textbf{bottom}) in the MSSM. Parameter scan ranges are shown in Eq. (\ref{scatter}), and other parameters  take the same values as Figure~\ref{fig:1}. The red (dark gray) circles denote the case of violating vacuum meta-stability and $m_{\tilde{\tau}1} > 0$ GeV. The gray circles  denote the case of satisfying vacuum meta-stability and $100$ GeV $> m_{\tilde{\tau}1} > 0$ GeV. The blue (black) circles denote the case of satisfying vacuum meta-stability and $ m_{\tilde{\tau}1} >100$ GeV. }
 \label{fig:3}
\end{figure}

For completeness, we probed the distribution of  $\Gamma (h \rightarrow \gamma \gamma ) / \Gamma (h \rightarrow \gamma \gamma )_{\text{SM}}$ by scanning the parameter space of the MSSM. Although there are many parameters in the MSSM, most of them (for example, $M_2$ and $M_A$) hardly influence the upper bound of  $\Gamma (h \rightarrow \gamma \gamma ) / \Gamma (h \rightarrow \gamma \gamma )_{\text{SM}}$ and we have checked this  numerically.
Therefore, we scanned only those parameters sensitive to   $\Gamma (h \rightarrow \gamma \gamma ) / \Gamma (h \rightarrow \gamma \gamma )_{\text{SM}}$ as follows, 
\beq
100 \textrm{ GeV} \leq \hspace{7pt} m_{\tilde{L}} \hspace{3pt}  &\leq & 1 \textrm{ TeV},\non
100 \textrm{ GeV} \leq \hspace{5pt}  m_{\tilde{\tau}R} &\leq &1 \textrm{ TeV}, \non
200 \textrm{ GeV} \leq \hspace{10pt} \mu \hspace{10pt}&  \leq & 1.5 \textrm{ TeV}, \non
20 \leq \hspace{2pt} \tan \beta  & \leq & 60 . \label{scatter}
\eeq
We considered a total of  one million points, where  all input parameters are value of the low energy scale. 
Scatter plots of the results are drawn in Figure \ref{fig:3}. We show  $\Gamma (h \rightarrow \gamma \gamma ) / \Gamma (h \rightarrow \gamma \gamma )_{\text{SM}}$ as a function of $\mu \tan\beta$ (left panel), $\tan \beta$ (right panel) and the lighter stau mass $m_{\tilde{\tau}1}$ (bottom panel) in the MSSM. The other parameters of Eq.~(\ref{scatter}) take the same values as Figure \ref{fig:1}. The red (dark gray) circles  denote the case of violating vacuum meta-stability and $m_{\tilde{\tau}1} > 0$ GeV. The gray circles denote the case of satisfying vacuum meta-stability and $100$ GeV $> m_{\tilde{\tau}1} > 0$ GeV. The blue  (black) circles denote the case of satisfying vacuum meta-stability and $ m_{\tilde{\tau}1} >100$ GeV.

In the panel  plotted as a function of $\mu \tan\beta$, the upper bound line of $\Gamma (h \rightarrow \gamma \gamma ) / \Gamma (h \rightarrow \gamma \gamma )_{\text{SM}}$ (the blue  (black) circles) reproduced the results of Figure \ref{fig:2}. Note that since we scanned  $m_{\tilde{L}},$ $m_{\tilde{\tau}R} < 1 $ TeV, the upper bound line of  $\Gamma (h \rightarrow \gamma \gamma ) / \Gamma (h \rightarrow \gamma \gamma )_{\text{SM}}$ could not be reproduced in the large $\mu \tan \beta $ region. 
In the panel plotted as a function of $\tan\beta$, we found that the upper bound of   $\Gamma (h \rightarrow \gamma \gamma ) / \Gamma (h \rightarrow \gamma \gamma )_{\text{SM}}$ is independent of $\tan \beta$. This implies that the upper bounds are determined only by the value of $\mu \tan \beta$. 
Finally, in the last panel (plotted as a function of $m_{\tilde{\tau}1}$), we found that  
  $\Gamma (h \rightarrow \gamma \gamma ) / \Gamma (h \rightarrow \gamma \gamma )_{\text{SM}}$ is  severely constrained by the vacuum stability in the light stau mass region.  
As a result, in the case of $m_{\tilde{\tau}1} =200$ GeV, $120$ GeV, $100$~GeV and $80$~GeV, the upper bounds of the enhancement are $9 \%$, $20 \%$, $25 \%$ and $40 \%$, respectively. 
Without considering  vacuum stability, in the case of $m_{\tilde{\tau}1} =100$ GeV, the upper bound of the enhancement is as large as $100 \%$ in the scanned parameters region (\ref{scatter}).

%\newpage
%%%%%%%%%%%%%%%%%%%%%%%%%%%%%%%%%%
%%%%%%%%%%%%%%%%%%%%%%%%%%%%%%%%%%

\section{Conclusions and Discussion}\label{sec5}
In this paper, motivated by recent observations of an enhanced diphoton Higgs decay and consistent $WW$ / $ZZ$ Higgs decay, we first analyzed the ratio $\Gamma (h \rightarrow \gamma \gamma ) / \Gamma (h \rightarrow \gamma \gamma ) _{SM}$ in a broad parameter region in the MSSM while applying  the stau vacuum meta-stability conditions. We found that the parameter regions of large enhancement to   $\Gamma (h \rightarrow \gamma \gamma ) / \Gamma (h \rightarrow \gamma \gamma ) _{SM}$  are severely constrained    from vacuum stability. We showed that  in the case of  the lighter stau mass  $m_{\tilde{\tau}1} =200$ GeV, $120$ GeV, $100$ GeV and $80$ GeV,  the upper bounds on the enhancement to the Higgs to diphoton rate are  $9 \%$, $20 \%$, $25 \%$ and $40 \%$, respectively. Especially, in the case of $m_{\tilde{\tau}1} = 100$ GeV, $\mu \tan \beta \sim 24$ TeV gives the largest enhancement to the Higgs to diphoton rate.

This result implies that if the stability of the scalar potential is taken into consideration, a large deviation between the  diphoton and $WW / ZZ$ signal strengths  can not be achieved in the MSSM.
Another implication is  that if we require a  $30\%$  enhancement to the diphoton signal strength in the MSSM,  $ \sigma (pp\rightarrow h)/\sigma (pp\rightarrow h)_{SM}  \times \Gamma (h \rightarrow \textrm{All})_{SM}/\Gamma (h \rightarrow \textrm{All})$ is required to be enhanced, and hence  an enhancement to the $WW/ZZ$ signal strength becomes inevitable.   

Note that the vacuum stability condition used in this paper arises from the tree-level  scalar potential \cite{Hisano:2010re}.
One should take into account the  vacuum stability condition which includes higher-order correction
since the parameter regions which can enhance the diphoton rate would receive large higher-order corrections. 
However, the estimation of the  vacuum stability condition which includes higher-order correction is complicated and 
the author reserves the study of the higher-order vacuum stability condition for future work.

Furthermore, note that not only in the MSSM but also in  general models, ``light charged scalar particles scenarios" or ``light charged vector-like lepton scenarios" would need large cubic or large quartic interaction with the Higgs boson in order to enhance the diphoton signal strength \cite{Carena:2012xa, ArkaniHamed:2012kq}. Large cubic scalar interaction, however, may suffer from vacuum instability just like the MSSM. Also, large quartic interaction may suffer from vacuum instability when dimensionless couplings become negative values, and from
Landau poles when dimensionless couplings rapidly blow up at high scales. Hence in the light charged scalar particle or vector-like lepton scenarios, the consideration of the vacuum stability and Landau pole is desirable.

At the end of this project, the Ref.~\cite{Sato:2012bf} was posted on arXiv.  The authors of this paper   also  first applied  the vacuum meta-stability to the diphoton signal strength in some gauge mediation models. As a result, they showed that the parameter regions which are consistent with the Higgs mass and muon $g-2$ can enhance  the $BR(h\rightarrow \gamma \gamma)/BR(h\rightarrow \gamma \gamma)_{SM}$ to $20 \%-30 \%$.

%%%%%%%%%%%%%%%%%%%%%%%%%%%%%%%%%%
%%%%%%%%%%%%%%%%%%%%%%%%%%%%%%%%%%
%%%%%%%%%%%%%%%%%%%%%%%%%%%%%%%%%%
%%%%%%%%%%%%%%%%%%%%%%%%%%%%%%%%%%
%%%%%                        Acknowledgements                    %%%%%%
\section*{Acknowledgements}
The author would like to thank Takeo Moroi for useful discussions, and also thank Koichi Hamaguchi and Motoi Endo for motive argument of this paper.

%%%%%%%%%%%%%%%%%%%%%%%%%%%%%%%%%%
%%%%%%%%%%%%%%%%%%%%%%%%%%%%%%%%%%
%%%%%%%%%%%%%%%%%%%%%%%%%%%%%%%%%%
%%%%%%%%%%%%%%%%%%%%%%%%%%%%%%%%%%
%%%%%                                 Appendix  A                         %%%%%%
\section*{Appendix}
\appendix
\section{Loop functions and Higgs couplings}\label{AppA}
The Loop functions $A_{i}^h (\tau)$ are given as follows
\beq
A_1^h (\tau) &=& 2 + 3 \tau + 3 \tau (2- \tau )  f(\tau) ,\non
A_{\frac{1}{2}}^h (\tau )&=& -2\tau  \left( 1+ (1 - \tau ) f(\tau ) \right) ,\non 
A_{0}^h(\tau )&=& \tau (1 - \tau f(\tau )),
\eeq
and
\beq
f(\tau) =\left\{
\begin{array}{l}
\arcsin ^2 (\sqrt{\frac{1}{\tau}}) ,\textrm{　　　　if $\tau \geq 1$,}\\
-\frac{1}{4} \left( \ln{ (\frac{\eta_{+}}{\eta_{-}})} - i \pi \right)^2,\textrm{　if $\tau \leq 1$,}
\end{array}
\right.
\eeq
where
\beq
\eta_{\pm} \equiv (1 \pm \sqrt{1 - \tau}).
\eeq

In the MSSM, the Higgs coupling constants are given as follows
 \beq
 g_{hWW}&=& \frac{g^2 v}{\sqrt{2}} \sin (\beta - \alpha) ,\\
 g_{hff(\textrm{up type})}&=& \frac{m_f}{\sqrt{2} v}\frac{\cos \alpha}{\sin \beta},\\
 g_{hff(\textrm{down type})}&=& \frac{m_f}{\sqrt{2} v}\frac{- \sin \alpha}{\cos \beta}, \\
 g_{h \tilde{f}_i  \tilde{f}_i (\textrm{up type})}&=& \left( (- I_{3,L} + (I_{3,L} + Y_L )\sin^2 \theta_W  ) \frac{g m_Z}{\cos \theta_W} \sin (\alpha + \beta )  + \frac{\sqrt{2} m_{f}^2 }{v} \frac{\cos \alpha}{\sin \beta} \right) (x^{f _i}_{L} )^2 \non
 & & + \left( - Y_R \sin^2 \theta_W   \frac{g m_Z}{\cos \theta_W}   \sin (\alpha + \beta )  + \frac{\sqrt{2} m_{f}^2 }{v} \frac{\cos \alpha}{\sin \beta} \right)  (x^{f _i}_{R} )^2 \non   
& & + \frac{\sqrt{2} m_{f}}{ v} \frac{\mu \sin \alpha + A_{f} \cos \alpha }{\sin \beta} x^{f _i}_{L} x^{f _i}_{R}   , \\
g_{h \tilde{f}_i  \tilde{f}_i (\textrm{down type})}&=& \left( (- I_{3,L} + (I_{3,L} + Y_L )\sin^2 \theta_W  ) \frac{g m_Z}{\cos \theta_W} \sin (\alpha + \beta )  - \frac{\sqrt{2} m_{f}^2 }{v} \frac{\sin \alpha}{\cos \beta} \right) (x^{f _i}_{L} )^2 \non
 & & + \left( - Y_R \sin^2 \theta_W   \frac{g m_Z}{\cos \theta_W}   \sin (\alpha + \beta )  - \frac{\sqrt{2} m_{f}^2 }{v} \frac{\sin \alpha}{\cos \beta} \right)  (x^{f _i}_{R} )^2 \non   
& & - \frac{\sqrt{2} m_{f}}{ v} \frac{\mu \cos \alpha + A_{f} \sin \alpha }{\cos \beta} x^{f _i}_{L} x^{f _i}_{R}   , \\
g_{h \chi^{+}_{i} \chi^{-}_{i}}&=& \frac{g }{\sqrt{2}} \left(-  \boldsymbol{V}_{i1} \boldsymbol{U}_{i2} \sin \alpha + \boldsymbol{V}_{i2 }\boldsymbol{U}_{i1} \cos \alpha \right) ,\\
g_{h H^{+} H^{-}}&=& g \left( m_W \sin (\beta - \alpha ) + \frac{m_Z \cos 2 \beta }{2 \cos \theta_W } \sin (\alpha + \beta ) \right),
 \eeq
 where $Y_{L/R}$ and $I_{3,L/R}$ are hypercharge and isospin of left/right-handed sfermion, sfermion mass eigenstates are $\tilde{f}_i = x^{f_i}_L  \tilde{f}_L + x^{f_i}_R  \tilde{f}_R$,  $\theta_W$ is the Weinberg angle, and  $\alpha$ is a rotation angle which translates the gauge-eigenstate basis CP-even Higgs mass matrix into the mass-eigenstate basis.  The chargino mass matrix is diagonalized to a real positive diagonal mass matrix by two $2 \times 2$ unitary matrices $\boldsymbol{U}$ and $\boldsymbol{V}$ as follows,
 \beq
 \boldsymbol{U}^{\ast} \begin{pmatrix} M_2 &\sqrt{2}m_W \sin \beta \\ \sqrt{2} m_W \cos \beta & \mu \end{pmatrix} \boldsymbol{V}^{\dag} = \begin{pmatrix} m_{ \chi^{\pm }_{1}} &0 \\ 0 &  m_{ \chi^{\pm }_{2}} \end{pmatrix} .
 \eeq 

%%%%%%%%%%%%%%%%%%%%%%%%%%%%%%%%%%
%%%%%%%%%%%%%%%%%%%%%%%%%%%%%%%%%%
%%%%%%%%%%%%%%%%%%%%%%%%%%%%%%%%%%
%%%%%%%%%%%%%%%%%%%%%%%%%%%%%%%%%%
%%%%%                                 Appendix  B                        %%%%%%

\section{Analysis of the global minimum of the scalar potential}\label{AppB}

In this appendix, we analyze numerically the global minimum of the scalar potential in four-field space ($h_d ,\  h_u, \ \tilde{L},\  \tilde{\tau}_R$), and probe  upper bound of $\mu$ requiring that the electroweak-breaking minimum is the global minimum. 
Note that $h_u$ is the same as $\phi$ in the main text and $h_d$ is the down-type neutral Higgs field.
The tree level scalar potential  which only includes dominant top/stop loop correction    in four-field space  is given as follows,
\beq
V &=& (\mu^2+ m_{H_d}^2 ) (H_d^0)^2+ (\mu^2 + m_{H_u}^2 ) (H_u^0)^2   - 2 B_{\mu}  H_d^0 H_u^0  \nonumber \\
&+& y_{\tau}^2 \left(  \left(  \tilde{L}^2  + \tilde{\tau}_R^2 \right) (H_d^0)^2 + \tilde{\tau}_R^2 \tilde{L}^2 \right) - 
 2 y_{\tau} \mu  \tilde{\tau}_R \tilde{L} H_u^0  + m_{\tilde{L}}^2 \tilde{L}^2 + m_{\tilde{\tau}_R}^2 \tilde{\tau}_R^2 
+ 2 y_{\tau} A_{\tau} \tilde{\tau}_R \tilde{L} H_d^0  \non
&+& \frac{g^{\prime 2}}{2 }\left( -\frac{1}{2}  (H_d^0)^2 + \frac{1}{2}(H_u^0)^2 - \frac{1}{2} \tilde{L}^2  +  \tilde{\tau}_R^2  \right)^2
+ \frac{g^2}{8}\left( (H_d^0)^2 -  ( H_u^0)^2 - \tilde{L}^2 \right)^2\non
&+& \frac{g^{\prime 2} + g^2 }{ 8} \Delta_t (H_u^0)^4 , \label{potential2}
\eeq
where $H_d^0 = v_1 + h_d /\sqrt{2}$, $H_u^0 = v_2 + h_u /\sqrt{2}$ and $\Delta_t $ is given by Eq.~(\ref{kore}). The scalar potential (\ref{potential2}) includes only real parts of scalar bosons. When the scalar potential (\ref{potential2}) is expanded around the electroweak-breaking vacuum and the down-type neutral Higgs field is omitted, it reproduces the scalar potential (\ref{potential}).

\begin{figure}[tbp]
 \begin{center}
 \includegraphics[width=76mm]{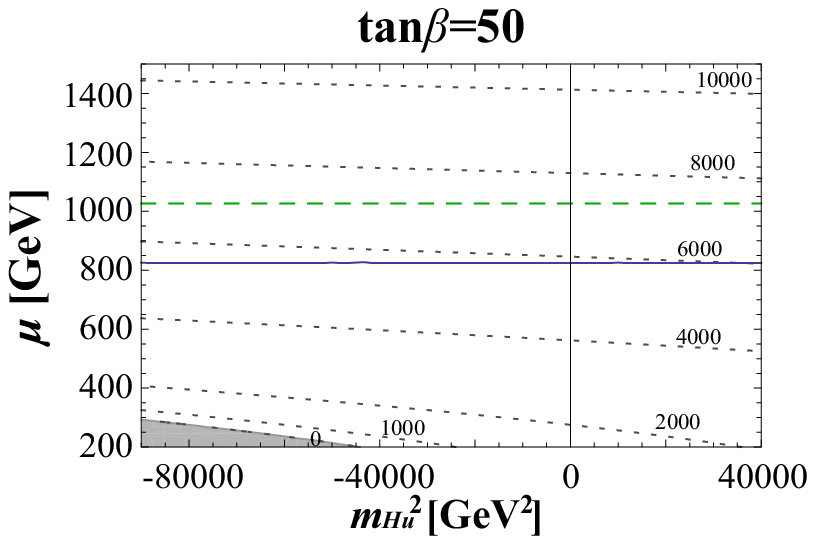} \textrm{} \includegraphics[width=76mm]{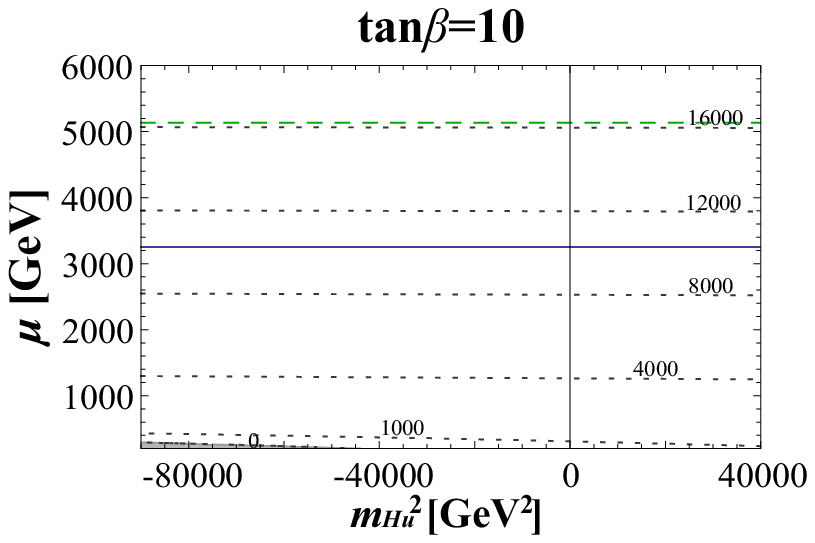} \\
 \end{center}
 \caption{The upper bound of $\mu$ requiring that the electroweak-breaking minimum is the global minimum as a function of $m_{H_u}^2$. We take $\tan \beta = 50$ (\textbf{left}) and $\tan \beta = 10$ (\textbf{right}). In both of the panels, $m_{\tilde{L}} = m_{\tilde{\tau}_R} = 400$ GeV , $A_{\tau} = 0$ GeV and $\Delta_t = 1$. Green dashed lines show the meta-stability bound (\ref{hisano}). Dotted lines are contour plots of $\sqrt{B_{\mu}}$ [GeV]. Gray areas denote the region where  electroweak symmetry breaking can not be achieved.}
 \label{fig:5}
\end{figure}

In Figure \ref{fig:5}, we show the upper bound of $\mu$ requiring that the electroweak-breaking minimum is the global minimum as a function of $m_{H_u}^2$  for $\tan \beta = 50$ (left-hand side) and $\tan \beta = 10$ (right-hand side). In both of the panels, $m_{\tilde{L}} = m_{\tilde{\tau}_R} = 400$~GeV, $A_{\tau} = 0$~GeV and $\Delta_t = 1$. Green dashed lines show the meta-stability bound (\ref{hisano}). Dotted lines are contour plots of $\sqrt{B_{\mu}}$. Gray areas denote the regions where electroweak symmetry breaking can not be achieved.

\begin{figure}[tbp]
 \begin{center}
 \includegraphics[width=80mm]{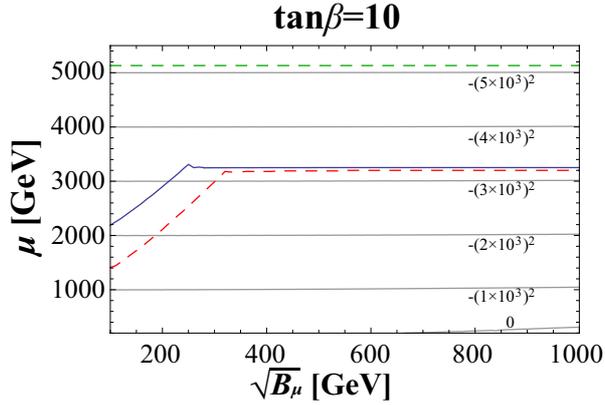}
 \end{center}
 \caption{The upper bound of $\mu$ requiring that the electroweak-breaking minimum is the global minimum as a function of $\sqrt{B_{\mu}}$. We take  $\tan \beta = 10$ and $A_{\tau} = 0$ GeV (blue solid line), $-500$ GeV (red dashed line). Green dashed line shows the meta-stability bound (\ref{hisano}). Gray lines are contour plots of $m_{H_u}^2$ [GeV$ ^2$].}
  \label{fig:6}
\end{figure}

Moreover in Figure \ref{fig:6}, we showed the upper bound of $\mu$ requiring that the electroweak-breaking minimum is the global minimum as a function of $\sqrt{B_{\mu}}$. We take $\tan \beta = 10$, $m_{\tilde{L}} = m_{\tilde{\tau}_R} = 400$ GeV, $\Delta_t = 1$ and $A_{\tau} = 0$ GeV (blue solid line), $-500$ GeV (red dashed line). Green dashed line shows the meta-stability bound (\ref{hisano}). Gray lines are contour plots of $m_{H_u}^2$.

We found that regardless of the sign of $m_{H_u}^2$, the parameter $m_{H_u}^2$ does not affect the upper bound of $\mu$.
We  also found that the low $\sqrt{B_{\mu}}$ and low $\tan \beta $ regions can affect the upper bound of $\mu$ and  these regions are sensitive to $A_{\tau}$. 
The reason is that because these regions  exhibit low $M_A$,  $h_d$ component gives considerable contribution in the scalar  potential (\ref{potential2}) , which then brings down another charged-breaking vacuum which becomes the global minimum earlier than the usual charged-breaking vacuum. 
Furthermore, we found that the shape of scalar potential (\ref{potential2}) remains unaltered by 
a change of $m_{H_u}^2$ and $\sqrt{B_{\mu}}$, except low $\sqrt{B_{\mu}}$ and low $\tan \beta $ regions.
These results imply that the approximate meta-stability conditional function (\ref{hisano}) can be considered reasonable except in the low $\sqrt{B_{\mu}}$ and low $\tan \beta $ regions.

\bibliography{ref}
%\end{flushleft}
 \end{document}